\documentstyle[preprint,eqsecnum,aps,epsf]{revtex}	

\newif\iftightenlines\tightenlinesfalse
\tightenlines\tightenlinestrue

\begin{document}
%

\hyphenation{mssm}
%
\preprint{\vbox{\baselineskip=14pt%
   \rightline{FSU-HEP-940625}\break 
   \rightline{UH-511-795-94}
}}
\title{MULTI-CHANNEL SEARCH FOR MINIMAL SUPERGRAVITY\\
AT $p\bar p$ and $e^+e^-$ COLLIDERS}
\author{Howard Baer$^1$, Chih-hao Chen$^1$, Ray Munroe$^1$,\\
Frank E. Paige$^2$ and Xerxes Tata$^3$}
\address{
$^1$Department of Physics,
Florida State University,
Tallahassee, FL 32306 USA
}
\address{
$^2$Brookhaven National Laboratory,
Upton, NY, 11973 USA
}
\address{
$^3$Department of Physics and Astronomy,
University of Hawaii,
Honolulu, HI 96822 USA
}
\date{\today}
\maketitle
\begin{abstract}

We examine the phenomenology of minimal supergravity models, assuming
only that the low energy theory has the minimal particle content, that
electroweak symmetry is radiatively broken, and that R-parity is
essentially conserved.  After delineating regions of supergravity
parameter space currently excluded by direct particle searches at LEP
and the Tevatron, we quantify how this search region will be expanded
when LEP~II and the Tevatron Main Injector upgrades become
operational. We describe how various experimental analyses can be
consistently combined within a single framework, resulting in a
multi-channel search for supersymmetry, but note that this analysis is
sensitive to specific assumptions about physics at the unification
scale.

\end{abstract}

\medskip



Grand Unified Theories (GUTs)\cite{ROSS} provide an attractive
synthesis of all gauge interactions. A striking prediction of these
models is that the proton is unstable. While the observed stability of
the proton can be understood if the GUT scale $M_X \agt 10^{16}$ GeV,
the origin and stability of the tiny ratio $M_W/M_X \sim 10^{-14}$
remain unexplained. Supersymmetry (SUSY)\cite{REV} provides an elegant
mechanism for the stability of the hierarchy between the two scales
provided that supersymmetric particles are lighter than $\sim 1$ TeV.
More recently, the realization\cite{UNIF} that the precision
measurements of the gauge couplings in experiments at LEP are
consistent with the simplest SUSY SU(5) GUT (with $M_{SUSY} \sim 1$
TeV) but {\it incompatible} with minimal non-SUSY SU(5) has motivated
many authors\cite{SPECTRA} to reexamine the expectations of sparticle
masses within the theoretically appealing, and relatively strongly
constrained, supergravity (SUGRA) framework. Studies of nucleon
decay\cite{PDK} as well as the collider phenomenology\cite{COLL} of
these models have also appeared.

Part of the appeal of these models lies in their economy. The masses and
couplings of all the sparticles are fixed in terms of just four
additional parameters, renormalized at some ultra-high scale at which
the physics is very simple. These parameters may be taken\cite{REV} to
be the common values of soft SUSY-breaking trilinear and bilinear
couplings ($A_0$ and $B_0$), a common SUSY-breaking scalar mass
($m_0$), and a common SUSY-breaking mass for the gauginos of the
unbroken grand unified group ($m_{1/2}$). For phenomenological
analyses, however, masses and couplings renormalized at the weak scale
are required. These can be readily obtained from the unification scale
parameters using renormalization group (RG) techniques\cite{RGE}. This
RG evolution leads\cite{REV} to calculable splittings between the
masses of the gluinos and the electroweak gauginos and between the
squarks and sleptons.  Unlike third generation sfermions, the first two
generations of squarks are approximately degenerate, consistent with
the absence of flavor-changing neutral currents in the K-meson
sector. A particular attraction of this framework is that electroweak
gauge symmetry is automatically broken down to electromagnetism when
the Higgs doublet masses are evolved down to the weak scale.

Without SUGRA unification, the Minimal Supersymmetric Standard Model
(MSSM) requires a plethora of input parameters, making experimental
analyses difficult.  While SUGRA GUTs indeed provide an economic
framework, we recognize that the assumptions about the physics at
ultra-high energy scales may prove to be incorrect and that the
unification of gauge couplings suggested by LEP experiments may turn
out to be a coincidence. Our approach here is to take this unification
seriously and to explore implications of SUGRA models for experiments
at the Tevatron and LEP~II. We shall assume that the low energy theory
has the minimal particle content, that electroweak symmetry is
radiatively broken, and that R-parity is essentially conserved. To
leave our analysis as general as possible\cite{IBAN}, however, we do
not commit to a particular GUT group\cite{FN1}, and hence we do not
include any constraints from nucleon decay or Yukawa coupling
unification. Also, we do not incorporate any fine-tuning constraints.
These constraints are somewhat subjective and only serve to
yield upper limits on sparticle masses.  Finally, we do not include
dark matter constraints since these can be simply evaded, {\it e.g.} by
allowing a small R-parity violation which would have no implication
for collider experiments.  Our purpose is to study the correlations
between various experimental searches for supersymmetic particles
arising from the fact that the masses and mixing patterns of all the
sparticles are determined by just four additional SUSY parameters. If
SUSY particles are ultimately discovered, a study of their properties
will allow us to test our underlying assumptions.

Toward this end, we have constructed a program to numerically solve
the RG equations (RGE) of the MSSM, implementing SUGRA relations
mentioned above as boundary conditions at the scale $M_X$ where the
running weak SU(2) and hypercharge gauge couplings come together. We
use the one-loop equations except for gauge couplings, for which we
include two-loop terms. We also include threshold corrections due to
various sparticle contributions entering the gauge coupling RGE's.
The weak scale mass parameters, $m$, are taken to be their running
values evaluated at the scale $Q=m$; {\it i.e.}, they are solutions of
$m=m(m)$. We convert the running masses of the top quark\cite{TAR} and
gluino\cite{MARTIN} to the corresponding physical (pole) masses. After
solving the RGE's, we use the evolved parameters to obtain the scalar
potential at the scale $M_Z$. As emphasized in Ref.\cite{GAMB}, the
results obtained using the tree-level potential are very sensitive to
the scale at which the potential is evaluated.  This situation is
ameliorated by including one loop corrections from the large top and
also bottom Yukawa interactions (these are important when $\tan\beta$
is large) to the potential. In practice, we use the two conditions
obtained by minimizing the one-loop effective potential with respect
to the Higgs fields to eliminate $B_0$ in favor of $\tan\beta$, the
ratio of the Higgs field vacuum expectation values, and to determine
(up to a sign) the superpotential Higgsino mass parameter, $\mu$, at
the weak scale.  Finally, the effective potential can also be used to
determine the pseudoscalar mass $m_{H_p}$, which, in turn, fixes the
other Higgs boson masses. The entire procedure is then iterated until
stable results are obtained.

The weak scale squark, slepton and gaugino masses, the $A$-parameters,
and the values of $\mu$, $m_{H_p}$, and $\tan\beta$ can be used as
inputs for ISAJET\cite{ISAJET} to compute the masses and mixing
patterns of charginos and neutralinos, third generation sfermions and
Higgs boson scalars, and their production cross sections and decay
patterns relevant for phenomenology. We have incorporated our
supergravity RGE solution into ISAJET 7.10. We suggest that a good way
to combine the results of different searches for SUSY is to express
them as limits in the ($m_0$--$m_{1/2}$) plane for several choices of
the other SUGRA parameters, {\it e.g.}, $\tan\beta = 2$, 10, and $A_0
= 0$, $\pm 2m_0$.  However, we urge the reader to keep in mind that the
framework depends on assumptions about physics at the scale $M_X$.
For instance, SUGRA models predict a specific form for the
squark-slepton mass splitting and the splitting amongst the various
squarks. Also, they generally predict that $|\mu|$ is considerably
larger than SUSY breaking electroweak gaugino masses, so that lighter
charginos and neutralinos are gaugino-like; this has important
implications\cite{BT} for sparticle decay patterns. We distinguish
such SUGRA assumptions from {\it e.g.} the assumed unification of
gaugino masses which follows from grand unification, and stress that
analyses\cite{OLD} with ``MSSM input parameters'' not correlated as in
SUGRA models are also necessary since these serve to illustrate the
sensitivity of the signals to unification scale assumptions, specific
to supergravity.

We now turn to a discussion of existing and expected SUSY limits in
the SUGRA framework as a function of $m_0$, $m_{1/2}$, $A_0$,
$\tan\beta$, $\mathop{\rm sgn}\mu$, and $m_t$.  Throughout this paper, we take
$m_t = 170$ GeV, as suggested by the recent\cite{TOP} analysis by the
CDF collaboration.  Our results, for $A_0 = 0$ (this does {\it not}
mean that the weak scale parameter $A_t$ vanishes) and $\tan\beta = 2$
are shown in the ($m_0$--$m_{1/2}$) plane in Fig.~1 for ({\it a}) $\mu
< 0$, and ({\it b}) $\mu > 0$. The physical gluino mass (for $m_0 =
500$ GeV) is also shown.  The shaded region is excluded by theoretical
considerations: in this region, the minimization conditions with the
correct value of $M_Z$ require $\mu^2 <0$ (so that the correct electroweak
symmetry breaking is not obtained), charge or color breaking minima
occur, or (as in the protrusion labeled LSP) the lightest supersymmetric
particle (LSP) is charged or colored (or the sneutrino), 
which is not allowed by
cosmology assuming the LSP is sufficiently long-lived. While this last
constraint could be avoided by allowing for some R-parity violation,
the region is also already excluded by experimental constraints as
discussed below.

The region below the solid lines in Fig.~1 is excluded by experiments
at LEP and the Tevatron. We include the following existing results:
\begin{enumerate}
\item Bounds\cite{LEP} on scalar particle masses from non-observation
of signals in $Z$ decays at LEP. Exclusive searches lead to a lower
limit of about $M_Z/2$ on the charged scalar masses while the
invisible width of the $Z$ boson results in a comparable bound on the
sneutrino mass. For clarity, we have only shown the boundary (labeled
$\tilde{\nu}$(43)) of the region where $m_{\tilde{\nu}} > 43$~GeV.
\item The LEP bound $m_{\widetilde{W}_1} > 47$~GeV on the mass of the lighter
chargino.
\item The bound $m_{H_{\ell}} \agt 60$ GeV on the mass of the lighter
Higgs boson. We have checked that the heavier Higgs boson masses
exceed 200--300~GeV along the boundary of this region, so that
$H_{\ell}$ is approximately the Standard Model (SM) Higgs boson so
that limits from experimental searches\cite{SMHIGGS} for $H_{SM}$,
are also approximately valid in the SUGRA scenario.
\item The region labeled $E\llap/_T$ recently excluded by the D0
Collaboration from the non-observation of an excess of $E\llap/_T$ events
from squark and gluino production at the Tevatron\cite{PATERNO}. To
obtain this boundary\cite{FN2}, we have converted the boundary of the
excluded region in the ($m_{\tilde{g}}$--$m_{\tilde{q}}$) plane shown in
Ref.\cite{PATERNO} to the corresponding boundary in the plane of
Fig.~1. For $\mu > 0$ this $E\llap/_T$ region would fall entirely within
the excluded region of LEP, and hence, is not shown.
\end{enumerate}
\noindent The cross-hatched line summarizes the boundary of the
parameter plane excluded by current constraints.

In the near future, the energy of LEP will be upgraded to beyond the
$WW$ threshold. Charged sparticles\cite{LEP2} and Higgs
bosons\cite{HIGGS2} with masses up to about 90~GeV should then be
detectable. Below the dashed-dotted lines labeled $H_{\ell}$(90) (in
Fig.~1{\it a}, $H_{\ell}$ is lighter than 90~GeV over the whole
plane), $\tilde{\ell}_R$(90) and $\widetilde{W}_1$(90), the corresponding particles are
lighter than 90~GeV, so that these lines roughly denote the boundary
of the parameter space that will be probed by direct sparticle
searches at LEP~II. After the main injector becomes operational, the
search for multijet plus $E\llap/_T$ or multilepton events at the Tevatron
should probe gluino masses up to about 250--300 GeV\cite{BKT},
depending on $m_{\tilde{q}}$. Also, Tevatron experiments should be sensitive
to clean high $p_T$ isolated trilepton signals from $\widetilde{W}_1\widetilde{Z}_2$
production\cite{BT,BAER,CHARGINO}. The sensitivity is larger for small
values of $m_0$ for which the sleptons are much lighter than squarks,
so that the leptonic decays of the neutralino, and sometimes, even of
the chargino are enhanced\cite{BT}.

The efficiency for detecting these trilepton events is sensitive to the
kinematics of the chargino and neutralino decays. We have used ISAJET
to perform a Monte Carlo study of this signal for various values of
($m_0, m_{1/2}$) in Fig.~1, using the cuts described in
Ref. \cite{BAER} appropriate for the Tevatron.  For $m_{1/2} = 100$
GeV, and for $m_0 = 100$, 150 and 500~GeV ({\it i.e.} well away from
the kinematic boundary for two-body decays of $\widetilde{W}_1$ or $\widetilde{Z}_2$),
which yield $m_{\widetilde{W}_1}\simeq m_{\widetilde{Z}_2}$ ($\simeq 2m_{\widetilde{Z}_1}$) = 94--100
GeV in Fig.~1{\it a}, we find an efficiency of $\sim 40$\%, consistent
with previous studies\cite{CHARGINO} of the signal. Except for the 
$m_0 = 100$ GeV case where it falls to 31\%,
this efficiency
increases by about 10\% when $m_{1/2}$ is increased to 140 GeV, but then
$\sigma$($\widetilde{W}_1\widetilde{Z}_2$) is considerably smaller. For 
($m_0,m_{1/2}$) = (20, 100) GeV where the charginos and
neutralinos decay via two-body modes, the efficiency is also $\sim 50$\%.
We have
also computed the efficiency for ($m_0,m_{1/2}$) = (80, 100) GeV, a
point very close to the dashed boundary, and find it to be only 1\%,
primarily due to the soft lepton from 
$\widetilde{W}_1\rightarrow\tilde{\nu}_L\ell$ decay, which has a branching
fraction of almost 90\%.
Finally, we have computed the efficiencies for $m_{1/2} =
120$ GeV and $m_0$ = (20,40,300) GeV in Fig.~1{\it b} where the
chargino and neutralino masses are somewhat smaller. For the $m_0 =20$
GeV case, we find an efficiency of about 30\%; however, for $m_0=40$ GeV,
it falls to just 2\%: this is because the
$Q$-value for the two body decay $\widetilde{Z}_2 \rightarrow\tilde{\ell}_R\ell$, which
has a branching fraction of 99\%, is only 5 GeV. For the $m_0=300$ GeV 
case the efficiency is $\sim 27$\%, somewhat smaller than 
the 3-body decay cases
of Fig. 1a.

Thus, for most sets of SUGRA parameters, the trilepton detection
efficiency ranges between 30--50\% depending on the parameters for
charginos and neutralino masses in the range 65--130 GeV.  By the end
of the current Tevatron run, it is expected that the CDF and D0
experiments will collectively accumulate a data sample exceeding 100
$pb^{-1}$. Since the physics backgrounds to the trilepton signal are
tiny\cite{BT,BAER,CHARGINO}, we expect that trilepton cross sections as
small as 200 $fb$, (corresponding to 5--10 events/100~$pb^{-1}$)
should lead to detectable signals.  The boundary of this region,
computed using ISAJET 7.10 with CTEQ2L\cite{CTEQ} structure functions,
is shown by the dashed line labeled $3\ell$(200 $fb$).  Since the
signal is essentially rate limited, we estimate that the trilepton
cross sections an order of magnitude smaller should be detectable
after the main injector commences operation. This expands the signal
region to below the dashed line labeled $3\ell$(20 $fb$).  It is
interesting to see that the trilepton signal may be observable for
chargino masses well in excess of the LEP~II reach (particularly for
smaller values of $m_0$) in keeping with previous
studies\cite{BT,BAER,CHARGINO}.  Finally, we note that the spoiler
modes $\widetilde{Z}_2\rightarrow\widetilde{Z}_1 H_{\ell}$ and $\widetilde{Z}_2\rightarrow\widetilde{Z}_1 Z$, which become
accessible above the horizontal dotted lines, do not significantly
limit the observability of the trilepton signal for the range of
parameters in this figure.

In order to illustrate how the phenomenology is altered when we change
$A_0$ and $\tan\beta$, we repeat this analysis in Fig.~2 for ({\it a})
$A_0 =-2m_0$, $\mu > 0$ and ({\it b}) $\tan\beta = 10$, $\mu < 0$,
with all other parameters as in Fig.~1.  The most obvious difference
in Fig.~2{\it a} is the large shaded region excluded by ``theory''.
In the large parabolic region for $m_0 > 150$ GeV, the large
mixing between the scalar top states induced by the large value of
$A_0$ results in $m_{\tilde{t}_1}^2 < 0$, so that the symmetry is
improperly broken.  In the slanted-line shaded region with small
values of $m_{1/2}$, $\tilde{t}_1$ is not tachyonic, but the parameter
$m_{\tilde{t}_R}^2 < 0$. While this may be physically allowed (as long as it
is positive at scales where electroweak symmetry remains unbroken also
in the Higgs sector), we caution ISAJET users that this causes
technical problems in the program. Fortunately, this region is
excluded by LEP constraints.  Below the dashed-dotted line labeled
$\tilde{t}_1$(90), the lighter of the two scalar top states should be
detectable either at LEP~II or at the Tevatron as discussed in
Ref.\cite{BST}. Also, there is a small sliver of the parameter space
(not shown) above the parabolic shaded region where the $\tilde{t}_1$ would
be the LSP, and another small region where the stop would have been
detectable at LEP. Fig.~2{\it a}, otherwise, resembles Fig.~1{\it b},
where $\mu$ is also positive, and except for the fact that the $\tilde{t}_1$
may also be experimentally accessible, the phenomenology is
qualitatively very similar in the two cases.  It should also be noted
that the $3\ell$ signal from $\widetilde{W}_1\widetilde{Z}_2$ production will be strongly
suppressed when the decay $\widetilde{W}_1\rightarrow\tilde{t}_1b$ becomes kinematically
accessible. 

Finally, for $\tan\beta =10$ and $A_0=0$ 
in Fig.~2{\it b}, the distinguishing feature is that
$\tilde{\tau_1}$, the lighter of the two $\tilde{\tau}$, can be significantly
lighter than the sleptons of the $e$ and $\mu$ families. This is
because of the $\tilde{\tau_L}$-$\tilde{\tau_R}$ mixing induced by the $\tau$ Yukawa
interactions, which are enhanced for the larger value of $\tan\beta$.
The region in the upper left corner is excluded since, there, the
$\tilde{\tau_1}$ would be the LSP. We also note that over the entire range of
parameters in Fig.~2{\it b} that are allowed by current experiments,
$H_{\ell}$ is too heavy to be discovered at LEP~II. It is also
interesting to see that the Higgs spoiler mode for the $3\ell$ signal
is never open in Fig.~2, although the $\widetilde{Z}_2 \rightarrow Z \widetilde{Z}_1$ spoiler opens
up near $m_{1/2} \sim 250$~GeV.

To summarize, we have examined collider phenomenology of minimal
supergravity GUTs with radiative electroweak symmetry breaking. Within
this framework, the masses and couplings of all the sparticles are
fixed in terms of just four additional SUSY parameters. As a result,
cross sections for various supersymmetric processes become correlated.
ISAJET~7.10 provides the option of incorporating SUGRA mass relations
and mixing patterns.  While SUSY search strategies for specific
processes remain essentially the same as in the MSSM framework,
constraints from different searches frequently complement one another
in the sense that they probe different regions of the SUGRA parameter
space.  For instance, the hatched line in Fig.~1{\it a}, which is
obtained by combining the sneutrino and chargino mass limits from LEP
experiments with the limit on the gluino mass from the Tevatron
denotes the boundary of the region currently excluded by sparticle
searches. Fig.~1 and Fig.~2 give a graphical summary of our results as
a function of SUGRA parameters. We see that the particular processes
which offer the largest reach for SUSY discovery are sensitive to all
these parameters. While the the search for sleptons at LEP~II and the
search for hadronically quiet trilepton events from $\widetilde{W}_1\widetilde{Z}_2$
production at the main injector upgrade of the Tevatron are
generically important probes, the search for charginos at LEP~II plays
an important role when $\mu > 0$ or when $\tan\beta$ is large.  On
the other hand, the search for scalar tops provides the potential for
exploring regions of parameters not accessible by other searches when
stop mixing effects are large (as in Fig.~2{\it a}).  It is also
interesting to see that over large regions of parameter space several
signals must simultaneously be present at the Tevatron and LEP~II.

Although, within this framework, the search for gluinos and squarks at
the Tevatron, which may probe gluino masses up to about 300~GeV, is
somewhat less competitive than other searches, we cannot overstress
the importance of direct $\tilde{g}$ and $\tilde{q}$ searches:  comparison of
$m_{\tilde{g}}$ with $m_{\widetilde{W}_1}$, or of $m_{\tilde{q}}$ with $m_{\tilde{\ell}}$ would
serve as crucial tests of the assumed unification of gaugino and
sfermion masses. In a similar vein, it is also important not to
abandon ``MSSM'' analyses where the values of $\mu$ and $m_{H_p}$ are
independently input, or the relation between different squark and
slepton masses relaxed. An optimal procedure may be to use SUGRA
models to provide the default values of ``MSSM'' input parameters, and
then to test the sensitivity of the resulting predictions on the
various SUGRA relations.  We conclude by noting that the observation
of sparticles would not only be a spectacular new discovery, but that
a measurement of their properties in experiments at the Tevatron and
LEP~II could serve as a window to physics at the unification scale.


\acknowledgments

We thank M. Paterno and A. White for motivation and discussions. We also
thank Manuel Drees for many fruitful conversations and for his help
in comparing our masses with his calculations.
This research was supported in part by the U.~S. Department of Energy
under grant numbers DE-FG-05-87ER40319,
DE-FG-03-94ER40833, and DE-AC02-76CH00016. HB was also supported by the
TNRLC SSC Fellowship program. 

%

%
\newpage


\ifpreprintsty
   \firstfigfalse	
\fi

\begin{figure}
\epsfysize=6.5in
\centerline{\epsfbox{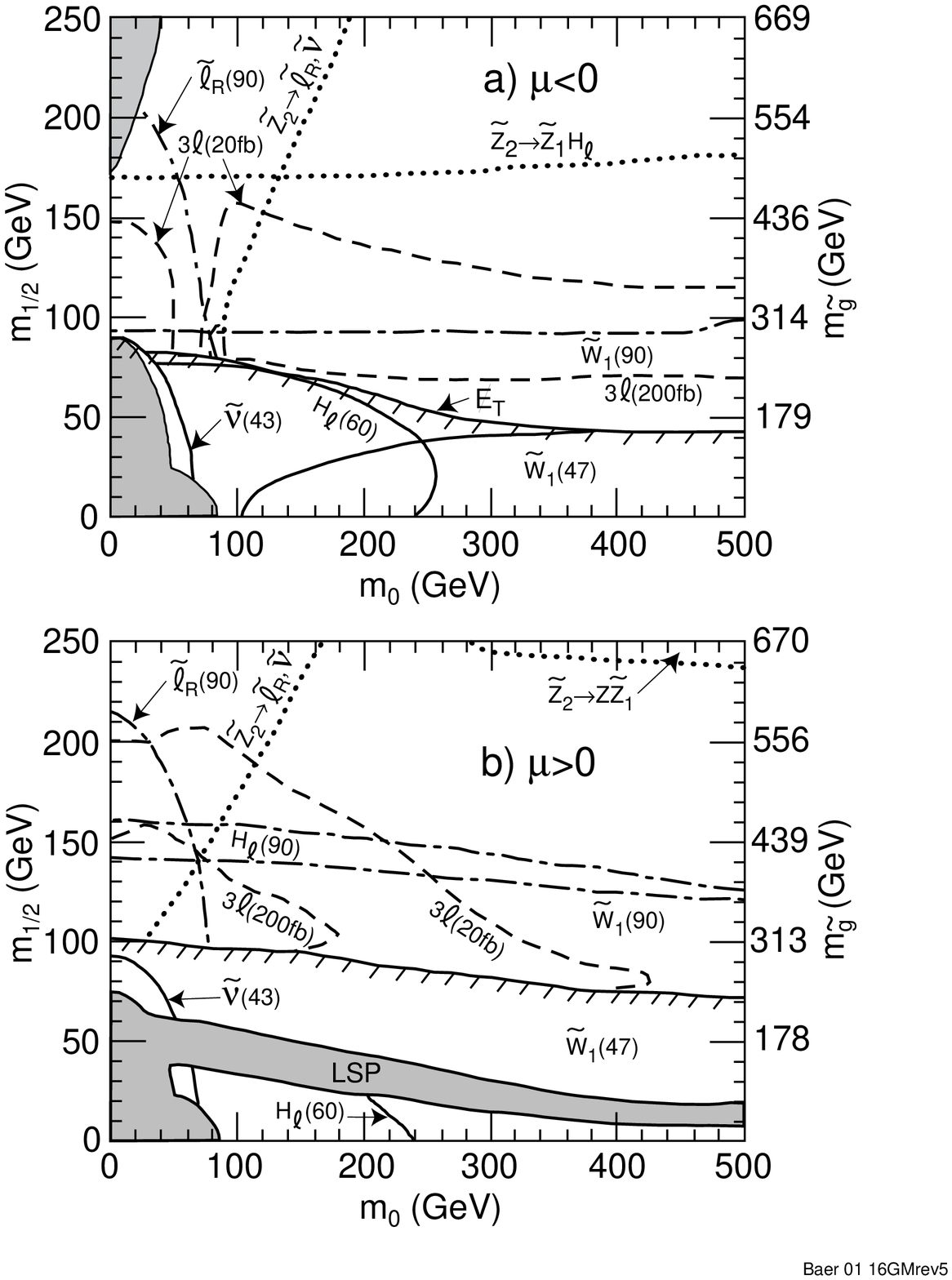}}
\medskip
\caption[]{Excluded regions and the reach of LEP II and the Tevatron
within the framework of the minimal supergravity model with
radiative electroweak symmetry breaking. We show the results in the 
$m_0-m_{1/2}$ plane for $A_0=0$, $\tan\beta=2$ and {\it a})$\mu < 0$ and 
{\it b})$\mu > 0$. The shaded region is excluded for theoretical reasons
discussed in the text, while the region below the hatched line is 
excluded by the experimental constraints from LEP and Tevatron. The area
below the dashed-dotted lines labeled $\tilde{\ell}_R(90)$, $\widetilde{W}_1(90)$ and
$H_{\ell}(90)$ [$m_{H_{\ell}}< 90$ GeV over the whole plane exhibited 
in case ({\it a})]
will be probed at LEP II via direct searches for sleptons, charginos, and
the lightest Higgs scalar,
respectively, while the region below the dashed lines should be accessible 
via the isolated trilepton search for $\widetilde{W}_1\widetilde{Z}_2$ events at the Tevatron
as discussed in the text. We terminate these dashed and dashed-dotted lines
at the hatched boundary for reasons of clarity.
Finally, the dotted lines denote the kinematic
boundary of various two body $\widetilde{Z}_2$ decays that significantly alter
the phenomenology.}
\end{figure}

\begin{figure}
\epsfysize=6.75in
\centerline{\epsfbox{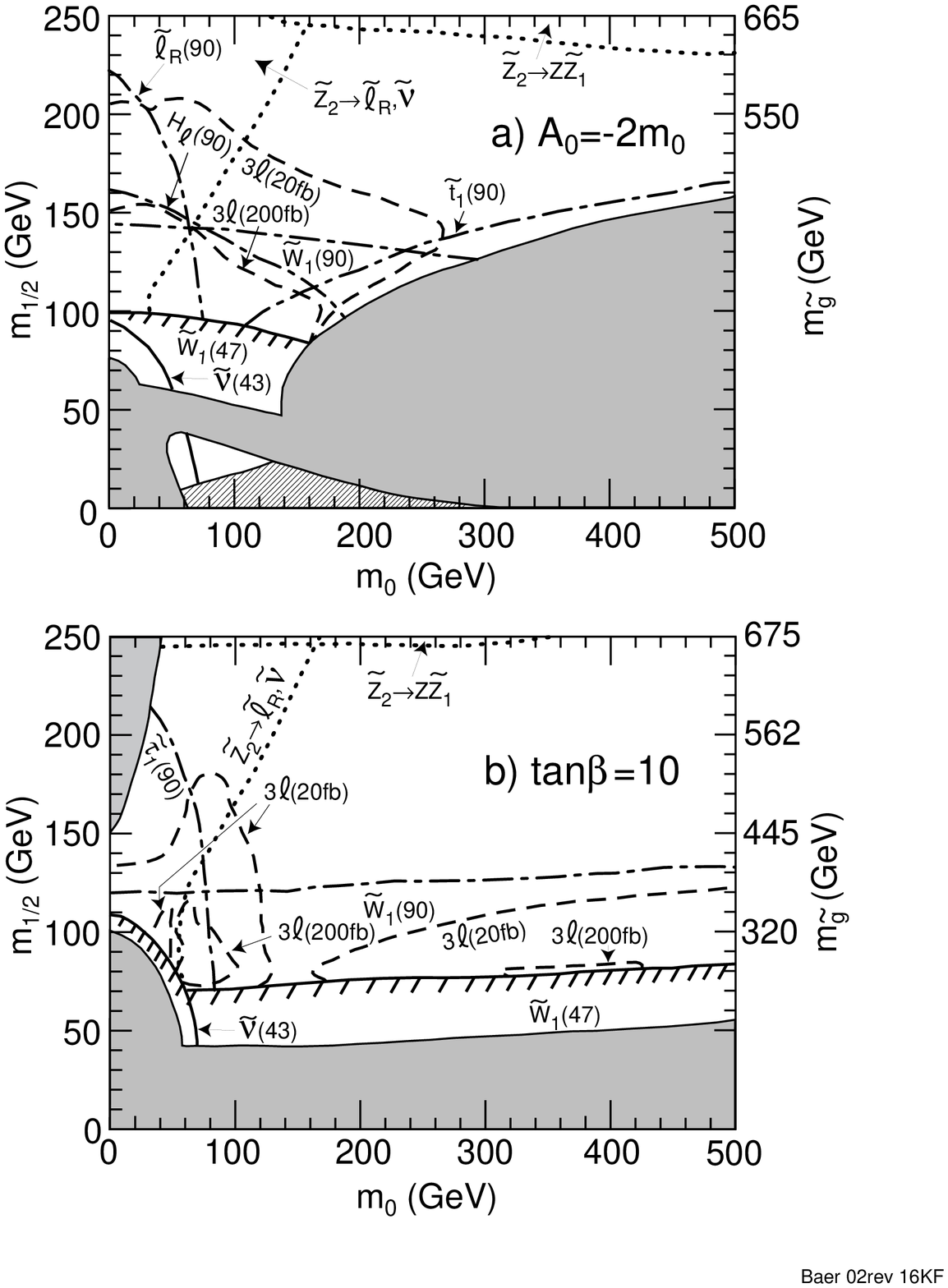}}
\medskip
\caption[]{The same as Fig.~1 except for {\it a})$A_0=-2m_0$, $\tan\beta=2$, 
$\mu > 0$ and 
{\it b})$\tan\beta = 10$, $A_0=0$, $\mu < 0$. 
The large shaded region in case~({\it a})
is excluded because $m_{\tilde{t}_1}^2 < 0$, while in the slanted line covered
region, $m_{\tilde{t}_R}^2 < 0$. In case ({\it b}), $H_{\ell}$ is too heavy to be
detected at LEP II, for parameter values not already excluded by experiment.}
\end{figure}


\begin{references}
\bibitem{FN1} We point out though that even the simplest supergravity SU(5)
GUT is completely consistent with all known experimental and cosmological
constraints (see Ref.\cite{COLL}) as also are flipped SU(5) $\times$ U(1)
models.
\end{references}
\end{document}